\documentclass[twocolumn,prd,preprintnumbers,superscriptaddress,landscape,nofootinbib]{revtex4}
\usepackage[utf8]{inputenc}
\usepackage{amsmath}
\usepackage{amsthm}
\usepackage{amssymb}
\usepackage{float}
\usepackage{braket}
\usepackage{graphicx}
\usepackage{url}
\usepackage{here}
\usepackage{natbib}
\usepackage{rotating}
\usepackage[colorlinks=true, pdfstartview=FitV, linkcolor=red, citecolor=blue, urlcolor=blue]{hyperref}
\usepackage{slashed}
\usepackage{multirow}
\usepackage[normalem]{ulem}
\graphicspath{{./figures/}}

\newcommand{\be}{\begin{equation}}      
\newcommand{\ee}{\end{equation}}      
\newcommand{\bea}{\begin{eqnarray}}      
\newcommand{\eea}{\end{eqnarray}}

\newcommand{\Tr}{\mathrm{Tr}}

\newcommand{\ctext}[1]{\raise0.2ex\hbox{\textcircled{\scriptsize{#1}}}}

\usepackage{tikz}
\usetikzlibrary{quantikz}

\theoremstyle{definition}

\theoremstyle{remark}

\usepackage[normalem]{ulem}

\begin{document}
\title{Demonstration of Quantum Energy Teleportation on Superconducting Quantum Hardware} 
\author{Kazuki Ikeda}
\email[]{kazuki7131@gmail.com}
\email[]{kazuki.ikeda@stonybrook.edu}
\affiliation{Co-design Center for Quantum Advantage, Stony Brook University, Stony Brook, New York 11794-3800, USA}
\affiliation{Center for Nuclear Theory, Department of Physics and Astronomy, Stony Brook University, Stony Brook, New York 11794-3800, USA}

\bibliographystyle{unsrt}

\begin{abstract}
Teleporting physical quantities to remote locations is a remaining key challenge for quantum information science and technology. Quantum teleportation has enabled the transfer of quantum information, but teleportation of quantum physical quantities has not yet been realized. Here we report the realization and observation of quantum energy teleportation on real superconducting quantum hardware. We achieve this by using several IBM’s superconducting quantum computers. The results are consistent with the exact solution of the theory and are improved by the mitigation of measurement error. Quantum energy teleportation requires only local operations and classical communication. Therefore our results provide a realistic benchmark that is fully achievable with current quantum computing and communication technologies.
\end{abstract}

\maketitle

\section{Quantum Energy Teleportation}
While it is fairly widely known that information about quantum states can be transported to remote locations~\cite{PhysRevLett.70.1895,furusawa1998unconditional,2015NaPho...9..641P,takeda2013deterministic}, it is less well known that quantum state energy can be similarly transmitted, despite its impact and potential for future applications. Quantum information transferred by quantum state teleportation (QST) is not a physical quantity, but energy is a distinct physical quantity. Transferring physical quantities to remote locations is an unexplored area of technology. Quantum Energy Teleportation (QET) was proposed by Hotta about 15 years ago and has been studied theoretically for spin chains~\cite{HOTTA20085671,2009JPSJ...78c4001H,2015JPhA...48q5302T},
an ion trap system~\cite{2009PhRvA..80d2323H},
a quantum Hall system~\cite{PhysRevA.84.032336}, and other various theoretical systems~ \cite{PhysRevA.82.042329,Hotta_2010}. QET has only recently been experimentally validated using an NMR setup~\cite{PhysRevLett.130.110801}. 

The purpose of this paper is to make the experimental verification of QET with actual cloud quantum computers in the most visible way, and to establish the optimized quantum circuits that make it possible. We achieved the realization of QET using some IBM superconducting quantum computers by applying quantum error mitigation~\cite{2018PhRvX...8c1027E,2022npjQI...8..114T,2022arXiv221000921C}. The quantum hardware we used includes IBM's quantum computer \texttt{ibmq\_lima}, which is available free of charge to everyone in the world. The quantum algorithm used in this work is open access to the public~\cite{Ikeda_Quantum_Energy_Teleportation_2023}, where quantum circuit implementation of QET is provided and real-time information to the latest machine properties is accessible. Using the quantum circuits provided in this paper, anyone will be able to reproduce the results and QETs of this study. Since all the properties of quantum computers are publicly available in real time, it will be possible for anyone to verify the QET protocol, regardless of whether one owns a quantum device or not. The methods we have established can be applied to any system capable of QET.

Although QET is conceptually similar to QST, here it is important to emphasize that, it is classical information, not energy, that is sent, and the intermediate subsystem along the channel between the sender (Alice) and receiver (Bob) is not excited by the energy carriers of the system during the short duration of the QET process. Thus, the time scale of energy transport by QET is much shorter than the time scale of heat generation in the natural time evolution. Bob can extract energy from a system by performing operations on his local system based on the classical information transmitted by Alice. By teleportation of energy or energy transfer, we mean that Bob can receive energy much faster (at the speed of light) than the energy that can be transmitted from Alice to Bob in the natural time evolution of the system.

In what follows, we explain that QET is a universal means of quantum energy extraction mediated by a many-body quantum system. Any non-trivial local operations, including measurements on the ground state of a quantum many-body system, give rise to excited states, which in turn increase the energy expectation value. Note that the increase in energy is supplied by the experimental devices. An important property of the ground state of a quantum many-body system is that it has entanglement, which brings about local quantum fluctuations in the global ground state. In QET, measurement plays an important role. Local measurement of the quantum state at a subsystem $A$ destroys this ground state entanglement. At the same time, energy $E_A$ from the device making the measurement is injected into the entire system. The injected energy $E_A$ stays around the subsystem $A$ in the very early stages of time evolution, but operations around $A$ alone cannot extract $E_A$ from the system. This is because information about $E_A$ is also stored in remote locations other than $A$ due to the entanglement that exists prior to the measurement~\cite{2010PhLA..374.3416H}. QET is the protocol that makes this possible by combing LOCC and conditional operations. Note that, because of the mid-circuit measurement, the time evolution of the post-measurement state is not unitary. Up to this point, no special assumptions about the quantum system have been used. The crucial property of QET is that it can be realized entirely by the general nature of the ground state of the entangled quantum many-body system and the universal fact of measurement. Using this ground state entanglement, the phase diagrams of many-body systems are reproduced by QET, where Alice and Bob's coordinates are fixed and teleported energy to Bob's local system reproduces the phase structure~\cite{2023arXiv230209630I,PhysRevD.107.L071502}. Those findings suggest that global structures such as symmetry, topology, and long-correlation of many-body systems can be detected by simple LOCC; one does not need to measure all qubits/states, which is advantageous for research in quantum physics using quantum computers. 

\begin{figure*}
    \centering
    \includegraphics[width=\linewidth]{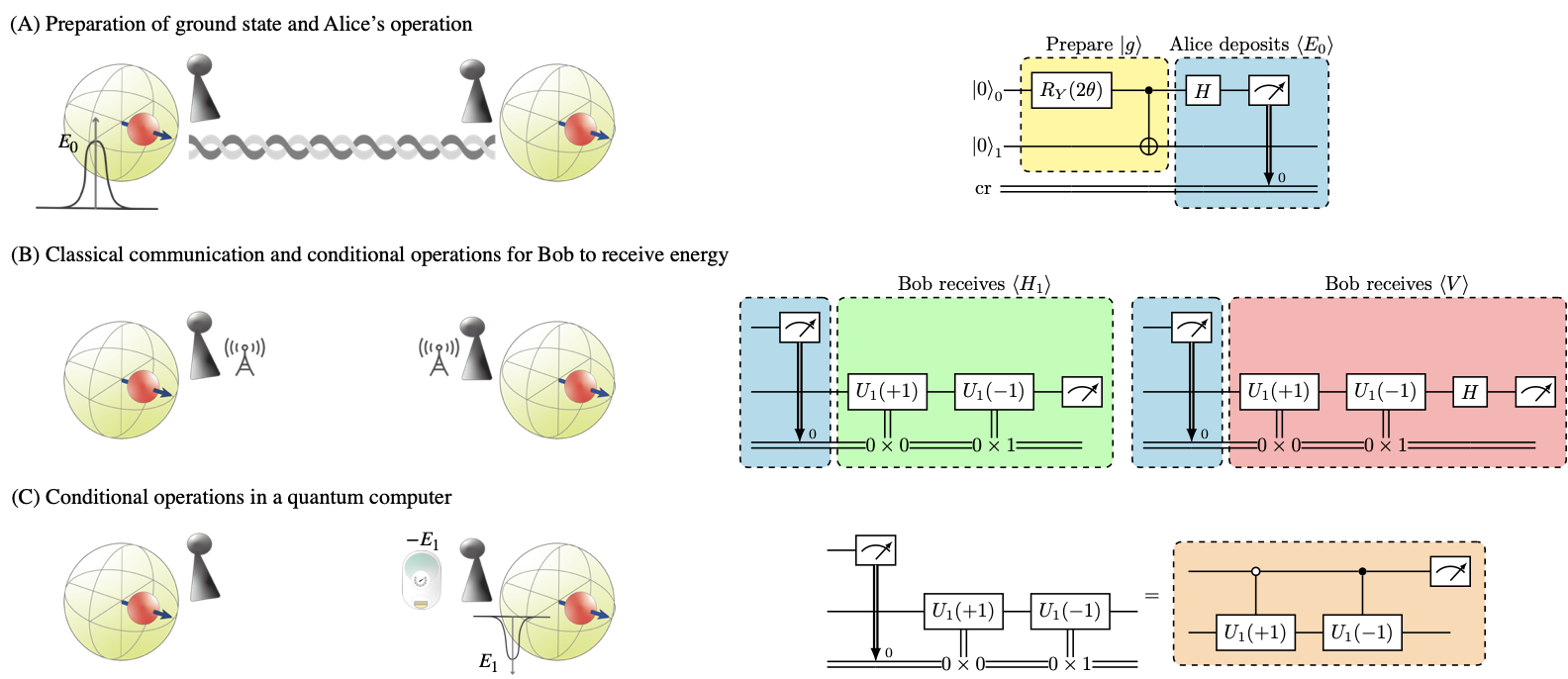}
    \caption{Quantum gate operations used for quantum energy teleportation. (A) preparation of ground state and Alice's $X_0$ measurement to deposit her energy. She tells Bob via classical communication whether $\mu=-1$ or $\mu=+1$ was observed.  (B) Bob's conditional operations to receive energy. He selects an operation $U_1(+1)$ or $U_1(-1)$ based on $\mu=+1$ or $-1$, corresponding to the Maxwell demon operation. (C) Equivalent implementation of Bob's operations on a quantum computer.}
    \label{fig:operation}
\end{figure*}

We work on the minimal QET model given in \cite{2011arXiv1101.3954H}. One of the purposes of this paper is to give a quantum circuit that utilizes QET with real quantum computers and quantum networks. The complete form of quantum circuits we used for QET is displayed in Fig.~\ref{fig:operation}, which is recently extended to long-range and large-scale quantum networks~\cite{2023arXiv230111884I}. The maximum circuit depth is 6 and the number of qubits used is 2. Hence, current quantum computers are powerful enough to implement QET. 

Let $k,h$ be positive real numbers. The Hamiltonian of the minimal model is
\begin{align}
    H_\text{tot}&=H_0+H_1+V,\\
    H_n&=hZ_n+\frac{h^2}{\sqrt{h^2+k^2}},~(n=0,1)\\
    V&=2kX_0X_1+\frac{2k^2}{\sqrt{h^2+k^2}}.
\end{align}
The ground state of $H_\text{tot}$ is 
\begin{equation}
\label{eq:groundstate}
    \ket{g}=\frac{1}{\sqrt{2}}\sqrt{1-\frac{h}{\sqrt{h^2+k^2}}}\ket{00}-\frac{1}{\sqrt{2}}\sqrt{1+\frac{h}{\sqrt{h^2+k^2}}}\ket{11},
\end{equation}

The constant terms in the Hamiltonians are added so that the ground state $\ket{g}$ of $H_\text{tot}$ returns the zero mean energy for all local and global Hamiltonians: 
\begin{equation}
\label{eq:ground}
\bra{g}H_\text{tot}\ket{g}=\bra{g}H_0\ket{g}=\bra{g}H_1\ket{g}=\bra{g}V\ket{g}=0. 
\end{equation}
However it should be noted that $\ket{g}$ is neither a ground state nor an eigenstate of $H_n,V, H_n+V~(n=0,1)$. The essence of QET is to extract negative ground state energy of those local and semi-local Hamiltonians. 

The QET protocol is as follows. First, Alice makes a measurement on her Pauli operator $X_0$ by $P_0(\mu)=\frac{1}{2}(1+\mu X_0)$ and then she obtains either $\mu=-1$ or $+1$. It turns out that Alice's expectation energy is $E_0=\frac{h^2}{\sqrt{h^2+k^2}}$.

Via a classical channel, Alice then sends her measurement result $\mu$ to Bob, who applies an operation $U_1(\mu)$ to his qubit and measures $H_1$ and $V$. She tells the result in a time $t$, which must be much shorter than the coupling time scale $t\ll 1/k$. In our experiment, $t=O(10)~ns$ and $1/k=O(100)~ns$. The density matrix $\rho_\text{QET}$ after Bob operates $U_1(\mu)$ to $P_0(\mu)\ket{g}$ is 
\begin{equation}
\label{eq:rho_QET}
    \rho_\text{QET}=\sum_{\mu\in\{-1,1\}}U_1(\mu)P_0(\mu)\ket{g}\bra{g}P_0(\mu)U^\dagger_1(\mu). 
\end{equation}
Using $\rho_\text{QET}$, the expected local energy at Bob's subsystem is evaluated as $\langle E_1\rangle=\Tr[\rho_\text{QET}(H_1+V)]$, which is negative in general. Due to the conservation of energy, $ E_B=-\langle E_1\rangle (>0)$ is extracted from the system by the device that operates $U_1(\mu)$~\cite{PhysRevD.78.045006}. In this way, Alice and Bob can transfer the energy of the quantum system by operations on their own local system and classical communication (LOCC).

\section{\label{sec:QET}Quantum Circuit Implementation of Quantum Energy Teleportation}
\subsection{\label{sec:setup}Preparation of Ground State}
The exact ground state $\ket{g}$ is prepared by 
\begin{equation}
\label{eq:gs_new}
    \ket{g}=\text{CNOT}(R_Y(2\theta)\otimes I)\ket{00},
\end{equation}
where $\theta=-\arccos\left(\frac{1}{\sqrt{2}}\sqrt{1-\frac{h}{\sqrt{h^2+k^2}}}\right)$. The corresponding quantum circuit is shown in Fig.~\ref{fig:operation}~(A).

\subsection{\label{sec:setp1}Step 1: Deposit Energy}
We use the following projective measurement operator  
\begin{equation}
    P_0(\mu)=\frac{1}{2}(1+\mu X_0).
\end{equation}
We measure Alice's $X$ operator, by which we obtain a state $\ket{+}$ or $\ket{-}$. This operation does not affect Bob's energy since $[X_0,V]=[X_0,H_1]=0$. Using $[P_0(\mu),V]=0$ and $\bra{+}Z\ket{+}=\bra{-}Z\ket{-}=0$, we find that Alice's mean energy to deposit is
\begin{equation}
    \langle E_0\rangle=\sum_{\mu\in\{-1,1\}}\bra{g}P_0(\mu)H_\text{tot}P_0(\mu)\ket{g}=\frac{h^2}{\sqrt{h^2+k^2}}.
\end{equation}

Alice's operation can be implemented on a quantum circuit in Fig~\ref{fig:operation} (A). $\langle E_0\rangle$ can be calculated with the output bit-strings $00,01,10,11$. Analytical values $\langle E_0\rangle$ and results with quantum computers for different pairs of $k$ and $h$ are summarized in Table \ref{tab:IBMQ_EA}.

\subsection{\label{sec:step2}Step 2: Receive Energy}
As soon as Alice observes $\mu\in\{-1,1\}$, she tells her result to Bob who operates $U_B(\mu)$ to his qubit and measures his energy. Here $U_B(\mu)$ is 
\begin{equation}
    U_1(\mu)=\cos\phi I-i\mu\sin\phi Y_1=R_Y(2\mu\phi),
\end{equation}
where $\phi$ obeys 
\begin{align}
    \cos(2\phi)&=\frac{h^2+2k^2}{\sqrt{(h^2+2k^2)^2+h^2k^2}}\\
    \sin(2\phi)&=\frac{hk}{\sqrt{(h^2+2k^2)^2+h^2k^2}}.
\end{align}
The average quantum state $\rho_\text{QET}$ eq.\eqref{eq:rho_QET} is obtained after Bob operates $U_1(\mu)$ to $P_0(\mu)\ket{g}$. Then the average energy Bob measures is 
\begin{equation}
\label{eq:QET}
    \langle E_1\rangle=\Tr[\rho_\text{QET}(H_1+V)]=\Tr[\rho_\text{QET}H_\text{tot}]-\langle E_0\rangle,
\end{equation}
where we used $[U_1(\mu),H_1]=0$. It is important that the map $\sum_{\mu\in\{-1,1\}}P_0(\mu)\ket{g}\bra{g}P_0(\mu)\to \rho_\text{QET}$ is not a unitary transformation. Therefore eq.~\eqref{eq:QET} can be negative. This is in contrast to eq.~\eqref{eq:Ene}.
     
Now let us explain quantum circuits for the QET protocol. Since $V$ and $H_1$ do not commute, measurement on those terms should be done separately. In other words, Bob measures $V$ and $H_1$ independently and obtains $\langle V\rangle$ and $\langle H_1\rangle$ statistically. As Fig.~\ref{fig:HV} in Appendix~\ref{sec:model} shows, $\langle V\rangle$ is always negative and $\langle H_1\rangle$ is always positive. Therefore it is sufficient for Bob to measure only $\langle V\rangle$ to receive energy by QET.

We consider $\langle V(\mu)\rangle=\bra{g} P_0(\mu)U^\dagger_1(\mu)VU_1(\mu)P_0(\mu)\ket{g}$. The quantum circuit to compute $\langle V(\mu)\rangle$ is shown in the right panel of Fig.~\ref{fig:operation} (B). It is important to note that, since Bob knows $\mu$ which contains Alice's information, he can obtain $\langle V\rangle=\Tr[\rho_\text{QET}V]$ by local measurement only, although $V$ is not a local operator. Similarly we can measure $H_1$ in $Z$-basis as in the left panel of Fig.~\ref{fig:operation} (B). The corresponding quantum circuit is obtained by removing the second Hadamard gate from the previous circuit Fig.~\ref{fig:operation}~ (C). On average the circuit generates the energy expectation value 
\begin{align}
\begin{aligned}
\label{eq:ana}
    \langle E_1\rangle&=\sum_{\mu\in\{-1,1\}}\bra{g}P_0(\mu)U^\dagger_1(\mu)(H_1+V)U_1(\mu)P_0(\mu)\ket{g}\\
    &=-\frac{1}{\sqrt{h^2+k^2}}[hk\sin(2\phi)-(h^2+2k^2)(1-\cos(2\phi))]. 
\end{aligned}
\end{align}
If $\phi$ is small, $\langle E_1\rangle$ is negative. Bob receives energy $\langle E_B\rangle=-\langle E_1\rangle$ on average. In Appendix~\ref{sec:Hotta}, we performed measurement of $V(\mu)$ and $H_1$ based on the quantum circuit Fig.~\ref{fig:operation} (B) and summarized data in Table~\ref{tab:simulator}, where numerical values are compared with analytical values given in eq.~\eqref{eq:ana}.

\subsection{\label{sec:real}QET on Real Quantum Hardware}
Here we describe how to implement conditional operations that may not be natively supported by many quantum computers and quantum devices. In the QET protocol, Bob's operation must be selected according to the results of Alice's measurements, as shown in Fig.~\ref{fig:operation} (B). Even in environments where conditional statements are not supported, QET can be implemented without problems through the technique of deferred measurement.

We can postpone Alice's measurement until the end of the circuit, and obtain the same results. The conditional operations can be created by a controlled $U$ gate $\Lambda(U)=\ket{0}\bra{0}\otimes I+\ket{1}\bra{1}\otimes U$ and an anti-controlled $U$ gate $(X\otimes I)\Lambda(U)(X\otimes I)$. One would find the equivalence between the following two circuits. We use the right circuit enclosed by the orange dashed frame in Fig.~\ref{fig:operation} (C).  

\begin{figure*}
    \centering
    \includegraphics[width=\linewidth]{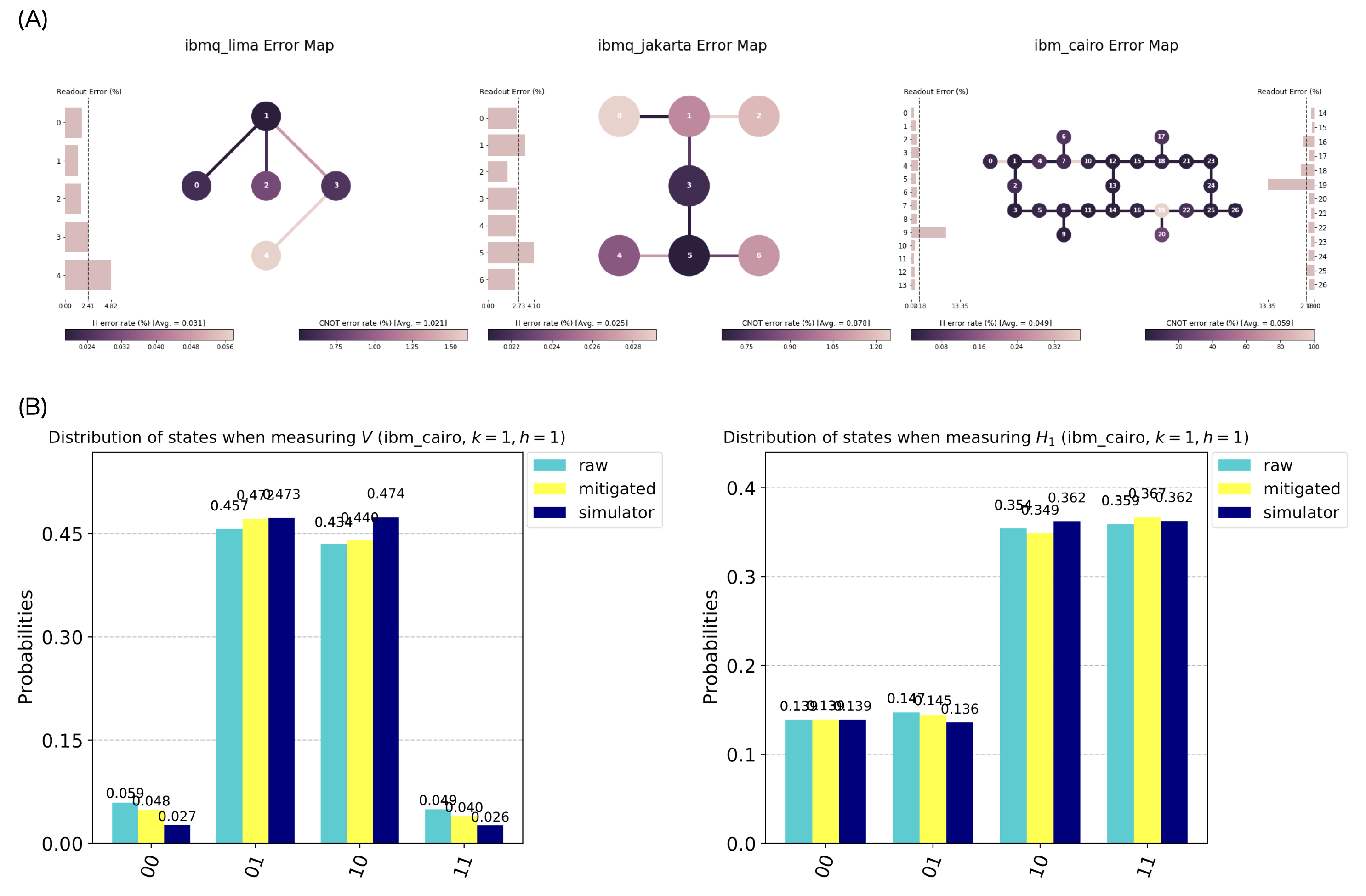}
    \caption{(A) properties of quantum computers we used. Each graph of qubits corresponds to the layout of the hardware. A direct CNOT gate can be applied to two qubits connected at the edge. (B) Distribution of states compared with a simulator \texttt{qasm\_simulator} and a quantum computer \texttt{ibm\_cairo} (raw results and mitigated results). Similar tendencies in the histograms were found for all other quantum computers used.}
    \label{fig:error_graph}
\end{figure*}

We performed quantum computation using 6 different types of IBM quantum hardware
 \texttt{ibmq\_lima}, \texttt{ibmq\_jakarta}, \texttt{ibmq\_hanoi}, \texttt{ibm\_cairo}, \texttt{ibm\_auckland} and \texttt{ibmq\_montreal}. The properties of each quantum computer can be seen from Fig.~\ref{fig:error_graph}. \texttt{ibmq\_lima} consists of 5 qubits (Fig.~\ref{fig:error_graph} [Left]) and \texttt{ibmq\_jakarta} has 7 qubits (Fig.~\ref{fig:error_graph} [Middle]). \texttt{ibm\_cairo} is a 27-qubit hardware, and \texttt{ibmq\_hanoi}, \texttt{ibm\_cairo}, \texttt{ibm\_auckland} and  \texttt{ibmq\_montreal} have the same graph structure as \texttt{ibm\_cairo}~(Fig.~\ref{fig:error_graph} [Right]). A direct CNOT gate can be applied to two qubits connected at the edge. We can choose two qubits placed on the graph of the hardware to perform quantum computation. We conducted the experiment by choosing two qubits connected at the edge with relatively small errors.

The time scale of QET can be estimated by comparing the gate time of a single-qubit gate and that of a two-qubit gate (CNOT gate). As shown in Table.~\ref{tab:processor}, the gate time of a CNOT gate is between 200-500 $ns$. On the other hand, the duration of a single qubit operator is between 20-40 $ns$~\footnote{For the latest real-time information of the duration of each operator in each machine, please visit~\cite{Ikeda_Quantum_Energy_Teleportation_2023}}. Therefore Bob can extract energy much faster than the time it takes for energy to be transferred from Alice to Bob in the natural unitary time evolution.
 
We also performed a simulation using a simulator \texttt{qasm\_simulator}, which can classically emulate gate operations on the same quantum circuits we used for quantum computation. We summarize results with \texttt{ibmq\_lima}, \texttt{ibmq\_jakarta} and \texttt{ibm\_cairo} in Table~\ref{tab:IBMQ_EA}. The results using the simulator agreed with the analytical solution with high accuracy, confirming that the quantum circuit was implemented correctly. More experimental results are summarized in Table~\ref{tab:complete} in  Appendix~\ref{sec:Add}. We describe details of machine properties and experimental conditions in Table \ref{tab:processor} in Appendix~\ref{sec:condition}. 
 
The most significant achievement in this study is the observation of negative energy $\langle E_1\rangle<0$. The value of $\langle V\rangle$ that was closest to the exact analysis value was -0.1079 ($h=1.5,k=1$ with \texttt{ibmq\_jakarta}), which is about 76\% accurate. As emphasised in Hotta's original works~\cite{HOTTA20085671,2009JPSJ...78c4001H,2015JPhA...48q5302T,2009PhRvA..80d2323H,PhysRevA.84.032336,PhysRevA.82.042329,Hotta_2010,2011arXiv1101.3954H}, after Alice observes her $X_0$, no unitary operation can make $\langle E_1\rangle$ negative (eq.~\eqref{eq:Ene}). In order for Bob to obtain the correct $\langle E_1\rangle$, Alice and Bob must repeat the experiment an enormous number of times, and the correct value of $\langle V\rangle$ and $\langle H_1\rangle$ can be obtained only when Alice and Bob communicate correctly in the quantum circuit in Fig.~\ref{fig:operation} (C). Distributions of states obtained by a quantum computer~\texttt{ibm\_cairo} are shown in Fig.~\ref{fig:error_graph} (B), where distributions of raw results and error mitigated results are compared with a simulator \texttt{qasm\_simulator}. We used a simple measurement error mitigation to determine the effects of measurement errors. We prepared a list of 4 measurement calibration circuits for the full Hilbert space. Then we immediately measured them to obtain the probability distributions. Then we applied the calibration matrix to correct the measured results. The average measurement fidelity when using each quantum computer is summarized in Table \ref{tab:processor} in Appendix~\ref{sec:condition}. The histograms of the observed states showed similar tendencies for all other quantum computers we used. It can be seen that the histograms obtained by the measurement of $H_1$ agree with the simulator results with good accuracy. The improvement of the values due to measurement error mitigation is also confirmed by the results in Table~\ref{tab:IBMQ_EA}. The observation of $V$ is of utmost importance in this study. Although the raw data from quantum computers deviated from the simulator results, in some cases, error mitigation improved them enough to observe negative energy expectation values.
 
It should also be emphasized that we observed negative $\langle V\rangle$ for all parameter $(k,h)$ combinations in all quantum computers used. As emphasized in Sec.~\ref{sec:step2}, the amount of energy available to Bob is greater if only V is observed since $\langle H_1\rangle$ is always positive (Fig.~\ref{fig:HV}). This would be enough for practical purposes. Note that the energy that Bob gains becomes smaller when he observes $H_1$. 

\begin{table*}
    \centering
    \begin{tabular}{|c|c|c|c|c|c|}\toprule
\cline{1-6}
Backend & &$(h,k)=(1,0.2)$&$(h,k)=(1,0.5)$& $(h,k)=(1,1)$ &$(h,k)=(1.5,1)$ \\\cline{1-6}
Analytical value&$\langle E_0\rangle$&0.9806&0.8944&0.7071&1.2481\\\hline
\texttt{qasm\_simulator}&&$0.9827\pm0.0031$&$0.8941\pm0.0001$&$0.7088\pm0.0001$& $1.2437\pm0.0047$\\\hline
\multicolumn{1}{ |c  }{\multirow{2}{*}{\texttt{ibmq\_lima}} } &
\multicolumn{1}{ |c| }{error mitigated} &$0.9423\pm0.0032$& $0.8169\pm0.0032$& $0.6560\pm0.0031$ &$1.2480\pm0.0047$    \\ \cline{2-6}
\multicolumn{1}{ |c  }{}                        &
\multicolumn{1}{ |c| }{unmitigated} &$0.9049\pm0.0017$&$0.8550\pm0.0032$&$0.6874\pm0.0031$&$1.4066\pm0.0047$    \\\hline
\multicolumn{1}{ |c  }{\multirow{2}{*}{\texttt{ibmq\_jakarta}} } &
\multicolumn{1}{ |c| }{error mitigated} &$0.9299\pm0.0056$& $0.8888\pm0.0056$& $ 0.7039\pm0.0056$ &$1.2318\pm0.0084$ \\ \cline{2-6}
\multicolumn{1}{ |c  }{}                        &
\multicolumn{1}{ |c| }{unmitigated} &$0.9542\pm0.0056$&$0.9089\pm0.0056$&$ 0.7232\pm0.0056$&$1.2624\pm0.0083$ \\ \hline
\multicolumn{1}{ |c  }{\multirow{2}{*}{\texttt{ibm\_cairo}} } &
\multicolumn{1}{ |c| }{error mitigated} &$0.9571\pm0.0032$& $0.8626\pm0.0031$& $0.7277\pm0.0031$ &$1.2072\pm0.0047$    \\ \cline{2-6}
\multicolumn{1}{ |c  }{}                        &
\multicolumn{1}{ |c| }{unmitigated} &$0.9578\pm0.0031$&$0.8735\pm0.0031$&$0.7362\pm0.0031$&$1.2236\pm0.0047$    \\ \cline{1-6}
\toprule
Analytical value&$\langle H_1\rangle$&0.0521&0.1873&0.2598 &0.3480\\\hline
\texttt{qasm\_simulator}&&$0.0547\pm0.0012$&$0.1857\pm0.0022$&$0.2550\pm0.0028$& $0.3487\pm0.0038$\\\hline
\multicolumn{1}{ |c  }{\multirow{2}{*}{\texttt{ibmq\_lima}} } &
\multicolumn{1}{ |c| }{error mitigated} &$0.0733\pm0.0032$& $0.1934\pm0.0032$& $0.2526\pm0.0032$ &$0.3590\pm0.0047$    \\ \cline{2-6}
\multicolumn{1}{ |c  }{}                        &
\multicolumn{1}{ |c| }{unmitigated} &$0.1295\pm0.0053$&$0.2422\pm0.0024$&$0.2949\pm0.0028$&$0.4302\pm0.0039$ \\ \cline{1-6}
\multicolumn{1}{ |c  }{\multirow{2}{*}{\texttt{ibmq\_jakarta}} }&
\multicolumn{1}{ |c| }{error mitigated} &$0.0736\pm0.0055$& $0.2018\pm0.0056$& $0.2491\pm0.0056$ &$0.3390\pm0.0084$ \\ \cline{2-6}
\multicolumn{1}{ |c  }{}                        &
\multicolumn{1}{ |c| }{unmitigated} &$0.0852\pm0.0022$&$0.2975\pm0.0045$&$ 0.3365\pm0.0052$&$0.4871\pm0.0073$\\ \hline
\multicolumn{1}{ |c  }{\multirow{2}{*}{\texttt{ibm\_cairo}} } &
\multicolumn{1}{ |c| }{error mitigated} &$0.0674\pm0.0032$& $0.1653\pm0.0031$& $0.2579\pm0.0031$ &$0.3559\pm0.0047$    \\ \cline{2-6}
\multicolumn{1}{ |c  }{}                        &
\multicolumn{1}{ |c| }{unmitigated} &$0.0905\pm0.0014$&$0.1825\pm0.0022$&$0.2630\pm0.0027$&$0.3737\pm0.0037$ \\
\toprule
Analytical value&$\langle V\rangle$&-0.0701&-0.2598 &-0.3746 &-0.4905\\\hline
\texttt{qasm\_simulator}&&$-0.0708\pm0.0012$&$-0.2608\pm0.0032$ &$-0.3729\pm0.0063$ &$-0.4921\pm0.0038$\\\hline
\multicolumn{1}{ |c  }{\multirow{2}{*}{\texttt{ibmq\_lima}} } &
\multicolumn{1}{ |c| }{error mitigated} &$-0.0655\pm0.0012$& $-0.2041\pm0.0031$& $-0.2744\pm0.0063$ &$-0.4091\pm0.0063$    \\ \cline{2-6}
\multicolumn{1}{ |c  }{}                        &
\multicolumn{1}{ |c| }{unmitigated} &$-0.0538\pm0.0011$&$-0.1471\pm0.0025$&$-0.1233\pm0.0041$&$-0.2737\pm0.0046$ \\ \cline{1-6}
\multicolumn{1}{ |c  }{\multirow{2}{*}{\texttt{ibmq\_jakarta}} } &
\multicolumn{1}{ |c| }{error mitigated} &$-0.0515\pm0.0022$& $-0.2348\pm0.0056$& $-0.3255\pm0.0112$ &$-0.4469\pm0.0112$ \\ \cline{2-6}
\multicolumn{1}{ |c  }{}                        &
\multicolumn{1}{ |c| }{unmitigated} &$-0.0338\pm0.0021$&$-0.1371\pm0.0046$&$ -0.0750\pm0.0075$&$-0.2229\pm0.0083$\\ \hline
\multicolumn{1}{ |c  }{\multirow{2}{*}{\texttt{ibm\_cairo}} } &
\multicolumn{1}{ |c| }{error mitigated} &$-0.0497\pm0.0013$& $-0.1968\pm0.0031$& $-0.2569\pm0.0063$ &$-0.3804\pm0.0063$    \\ \cline{2-6}
\multicolumn{1}{ |c  }{}                        &
\multicolumn{1}{ |c| }{unmitigated} &$-0.0471\pm0.0012$&$-0.1682\pm0.0026$&$-0.1733\pm0.0038$&$-0.3089\pm0.0045$    \\
\toprule
Analytical value&$\langle E_1\rangle$&-0.0180&-0.0726& -0.1147&-0.1425\\\hline
\texttt{qasm\_simulator}&&$-0.0161\pm0.0017$&$-0.0751\pm0.0039$8 &$-0.1179\pm0.0069$ &$-0.1433\pm0.0054$\\\hline
\multicolumn{1}{ |c  }{\multirow{2}{*}{\texttt{ibmq\_lima}} } &
\multicolumn{1}{ |c| }{error mitigated} &$0.0078\pm0.0034$& $-0.0107\pm0.0045$& $-0.0217\pm0.0071$ &$-0.0501\pm0.0079$    \\ \cline{2-6}
\multicolumn{1}{ |c  }{}                        &
\multicolumn{1}{ |c| }{unmitigated} &$0.0757\pm0.0054$&$0.0950\pm0.0035$&$0.1715\pm0.0050$&$0.1565\pm0.0060$ \\ \cline{1-6}
\multicolumn{1}{ |c  }{\multirow{2}{*}{\texttt{ibmq\_jakarta}} } &
\multicolumn{1}{ |c| }{error mitigated} &$0.0221\pm0.0059$& $-0.0330\pm0.0079$& $-0.0764\pm0.0125$ &$-0.1079\pm0.0140$ \\ \cline{2-6}
\multicolumn{1}{ |c  }{}                        &
\multicolumn{1}{ |c| }{unmitigated} &$0.0514\pm0.0030$&$0.1604\pm0.0064$&$ 0.2615\pm0.0091$&$0.2642\pm0.00111$\\ \cline{1-6}
\multicolumn{1}{ |c  }{\multirow{2}{*}{\texttt{ibm\_cairo}} } &
\multicolumn{1}{ |c| }{error mitigated} &$0.0177\pm0.0035$& $-0.0315\pm0.0044$& $0.0010\pm0.0070$ &$-0.0245\pm0.0079$    \\ \cline{2-6}
\multicolumn{1}{ |c  }{}                        &
\multicolumn{1}{ |c| }{unmitigated} &$0.0433\pm0.0018$&$0.0143\pm0.0034$&$0.0897\pm0.0047$&$0.0648\pm0.0058$    \\ \cline{1-6}
\end{tabular}
    \caption{Comparison between analytical values of $\langle E_0\rangle,\langle H_1\rangle, \langle V\rangle, \langle E_1\rangle$ and results from IBM's real quantum computers, \texttt{ibmq\_lima}, \texttt{ibmq\_jakarta} and \texttt{ibm\_cairo}. We evaluate $\langle E_1\rangle=\langle H_1\rangle+\langle V\rangle$. "error mitigated" means results using measurement error mitigation and "unmitigated" corresponds to results without measurement error mitigation. 
    }
    \label{tab:IBMQ_EA}
\end{table*}

\section{Implications for our real world}
Our results provide implications for new quantum communication technologies with respect to different phases in the short, medium and long term. It is important to note that, like quantum teleportation, energy can also be teleported only by LOCC. Reproducing the minimal QET model we used in our demonstration in a laboratory system is something that can be tackled in the short term with current quantum computing and communication technology. A quantum device with 2 qubits and a gate depth of 6 would be ready for immediate experiments. This is expected to lead to new developments in the use of quantum memory~\cite{doi:10.1126/sciadv.1600911,PhysRevApplied.18.064039,specht2011single}. Furthermore, verifying QET in a variety of quantum systems and materials beyond the minimal model is an important challenge for future applications.

QET without limit of distance is also provided~\cite{2014PhRvA..89a2311H}. The combination of QST and QET has been generalized to a simple universal QET protocol on arbitrarily large-scale quantum networks~\cite {2023arXiv230111884I}. The ability to transfer quantum energy over long distances at the speed of light will bring about a new revolution in quantum communication technology. For example, there is a long-distance ($\sim$158km) SBU/BNL quantum network in Long Island, New York~\cite{2021arXiv210112742D}. Various quantum networks have been developed~\cite{kimble2008quantum,chen2021integrated,doi:10.1126/science.abg1919}. Realizing QET on a quantum network, which is expected to be in practical use around the 2030s, would be a milestone toward realizing QET on a worldwide quantum network. 

The realization of a long-range QET will have important implications beyond the development of information and communication technology and quantum physics. More recently QET has been applied to quantum interactive proof and quantum cryptography on a large scale quantum network~\cite{Ikeda:2023yhm}, and it provides a secure authentication system with zero-knowledge. In addition, QET will help develop the quantum economy market. Information and energy are physical, but also economic. Allowing physical quantities to be traded concretely on the quantum network means that a new economic market will be born~\cite{ikeda2022theory}. Quantum teleportation is an established technology and is being developed for practical use. In addition to this, if QET is put to practical use, it will mean that various quantum resources will be at the disposal of us. The expected value of the Hermite operator is called energy, but it need not literally be used only as energy. Teleported energy can be used as energy, as well as for other uses. The ability to teleport a concrete physical quantity, energy, means that quantum information will have added value. In a quantum market where Alice, Bob, and Charlie exist, if Bob can get more energy from Charlie than from Alice, Bob may prefer to do business with Charlie rather than Alice, and he may prefer an entangle state with Charlie. However, depending on transaction costs, Bob may choose Alice. A lot of such game-theoretic situations can be created~\cite{2020QuIP...19...25I,Ikeda2023,2021QuIP...20..387I,2022arXiv221102073I,2021QuIP...20..313I}. This implies that quantum information economics (which does not yet exist) will become a meaningful idea in the future.

\section*{Acknowledgement}
I thank David Frenklakh, Adrien Florio, Sebastian Grieninger, Fangcheng He, Dmitri Kharzeev, Yuta Kikuchi, Vladimir Korepin, Qiang Li, Adam Lowe, Shuzhe Shi, Hiroki Sukeno, Tzu-Chieh Wei, Kwangmin Yu and Ismail Zahed for fruitful communication and collaboration. After posting to the first version of this paper to arXiv on 7 Jan 2023, Martín-Martíne told me about his demonstration with an NMR. I  thank him for the communication. I thank Megumi Ikeda for providing the cartoons. I acknowledge the use of IBM quantum computers. I was supported by the U.S. Department of Energy, Office of Science, National Quantum Information Science Research Centers, Co-design Center for Quantum Advantage (C2QA) under Contract No.DESC0012704. 

\section*{References}
\bibliographystyle{utphys}
\bibliography{ref}

\clearpage

\appendix
\section{\label{sec:model}Description of the Model}
\subsection{Quantum Gates and Measurement}
Here we give a self-contained description of the background knowledge of the text. We use the following one-qubit operators whose matrix representations are given as 
\begin{equation}
\begin{aligned}
    X&=\begin{pmatrix}
        0&1\\
        1&0
    \end{pmatrix},
    Y=\begin{pmatrix}
        0&-i\\
        i&0
    \end{pmatrix},\\
    Z&=\begin{pmatrix}
        1&0\\
        0&-1
    \end{pmatrix},
    H=\frac{1}{\sqrt{2}}\begin{pmatrix}
        1&1\\
        1&-1
    \end{pmatrix}.
\end{aligned}
\end{equation}
We use $\ket{0}=\binom{1}{0},\ket{1}=\binom{0}{1}$ for the computational basis states, which are eigenstates of $Z$: $Z\ket{0}=\ket{0},Z\ket{1}=-\ket{1}$. We also work with another basis vectors $\ket{\pm}=\frac{\ket{0}\pm\ket{1}}{\sqrt{2}}$. They are eignestates of $X$: $X\ket{-}=-\ket{-},X\ket{+}=-\ket{+}$. Note that $\ket{\pm}$ are created by applying $H$ to $\ket{0}$ and $\ket{1}$; $H\ket{0}=\ket{+},H\ket{1}=\ket{-}$. Those are used for measuring $H_{n},V~(n=1,2)$ in the QET protocol. For example, Alice finds $\mu=\pm1$ by observing the eigenvalues $\pm1$ of her local Pauli $X$ operator. 

The rotation of $X,Y,Z$ is defined by 
\begin{equation}
    R_X(\alpha)=e^{-i\frac{\alpha}{2} X},~R_Y(\alpha)=e^{-i\frac{\alpha}{2} Y},~R_Z(\alpha)=e^{-i\frac{\alpha}{2} Z}.
\end{equation}

We use two-qubit gate operations. In general, a control $U$ operation $\Lambda(U)$ is defined by 
\begin{equation}
    \Lambda(U)=\ket{0}\bra{0}\otimes I+\ket{1}\bra{1}\otimes U
\end{equation}
and the corresponding diagram is drwan as 
\begin{figure}[H]
    \centering
\begin{quantikz}
\midstick[2,brackets=none]{control $U$=}\qw&\ctrl{1}&\qw\\
\qw&\gate{U}&\qw
\end{quantikz}
\end{figure}

One of the most frequently used controlled gates is a CNOT gate $\text{CNOT}=\Lambda(X)$, whose diagram is especially drawn as
\begin{figure}[H]
    \centering
\begin{quantikz}
\midstick[2,brackets=none]{CNOT=}\qw&\ctrl{1}&\qw\\
\qw&\targ{}&\qw
\end{quantikz}
\end{figure}
Using the formula $\text{CNOT}(a\ket{0}+b\ket{1})\ket{0}=a\ket{00}+b\ket{11}$, it is easy to check 
\begin{equation}
    \text{CNOT}(R_Y(2\theta)\otimes I)\ket{00}=\cos\theta\ket{00}+\sin\theta\ket{11}. 
\end{equation}
When $\theta=-\arccos\left(\frac{1}{\sqrt{2}}\sqrt{1-\frac{h}{\sqrt{h^2+k^2}}}\right)$, we find that eq.~\eqref{eq:gs_new} is indeed the ground state $\ket{g}$. 

It is convenient to define an anti-control gate, which is activated when the control bit is in state $\ket{0}$: $\ket{1}\bra{1}\otimes I+\ket{0}\bra{0}\otimes U$, whose diagram is drawn as  
\begin{figure}[H]
    \centering
\begin{quantikz}
\midstick[2,brackets=none]{Anti-control $U$=}\qw&\octrl{1}&\qw\\
\qw&\gate{U}&\qw
\end{quantikz}
=
\begin{quantikz}
\gate{X}&\ctrl{1}&\gate{X}\\
\qw&\gate{U}&\qw
\end{quantikz}
\end{figure}

Now we describe the measurement of quantum operators. We measure $Z_{1}$ and $X_0X_1$. Measurement of $Z_1$ is done by the following circuit
\begin{figure}[H]
    \centering
    \begin{quantikz}
&\meter{}\\
&\meter{}
\end{quantikz}
\end{figure}
The output of the measurement is a bit string $b_0b_1\in\{00,01,10,11\}$. Since the eigenvalues of $Z$ are $-1,1$, we convert the bit string into $1-2b_1$. Let $n_\text{shot}$ be the number of repetitions of the circuit, and $\text{counts}_{b_0b_1}$ be the number of times $b_0$ and $b_1$ are detected. Therefore $\frac{\text{counts}_{b_0b_{1}}}{n_\text{shots}}$ is the probability that a bit string $b_0b_1$ is obtained. Then the expectation value of $Z_1$ is computed by the formula
\begin{align}
 \langle Z_1\rangle&=\sum_{b_0,b_1}(1-2b_1)\frac{\text{counts}_{b_0b_{1}}}{n_\text{shots}}.
\end{align}

Measurement of $X_0X_1$ is done by the following circuit
\begin{figure}[H]
    \centering
    \begin{quantikz}
&\gate{H}&\meter{}\\
&\gate{H}&\meter{}
\end{quantikz}
\end{figure}
As we described previously, $H$ maps $\ket{0},\ket{1}$ to $\ket{+},\ket{-}$, which are eigenvectors of $X$. The output of the measurement is again a bit string $b_0b_1\in\{00,01,10,11\}$. They are converted to the eigenvalues of $X_0X_1$ by $(1-2b_0)(1-2b_1)$. Then the expectation value of $X_0X_1$ is computed by the formula
\begin{align}
 \langle X_0X_1\rangle&=\sum_{b_0,b_1}(1-2b_0)(1-2b_1)\frac{\text{counts}_{b_0b_{1}}}{n_\text{shots}}.
\end{align}

\subsection{\label{sec:Model_QET} Some details of the model}
\begin{figure*}
    \centering
    \includegraphics[width=\linewidth]{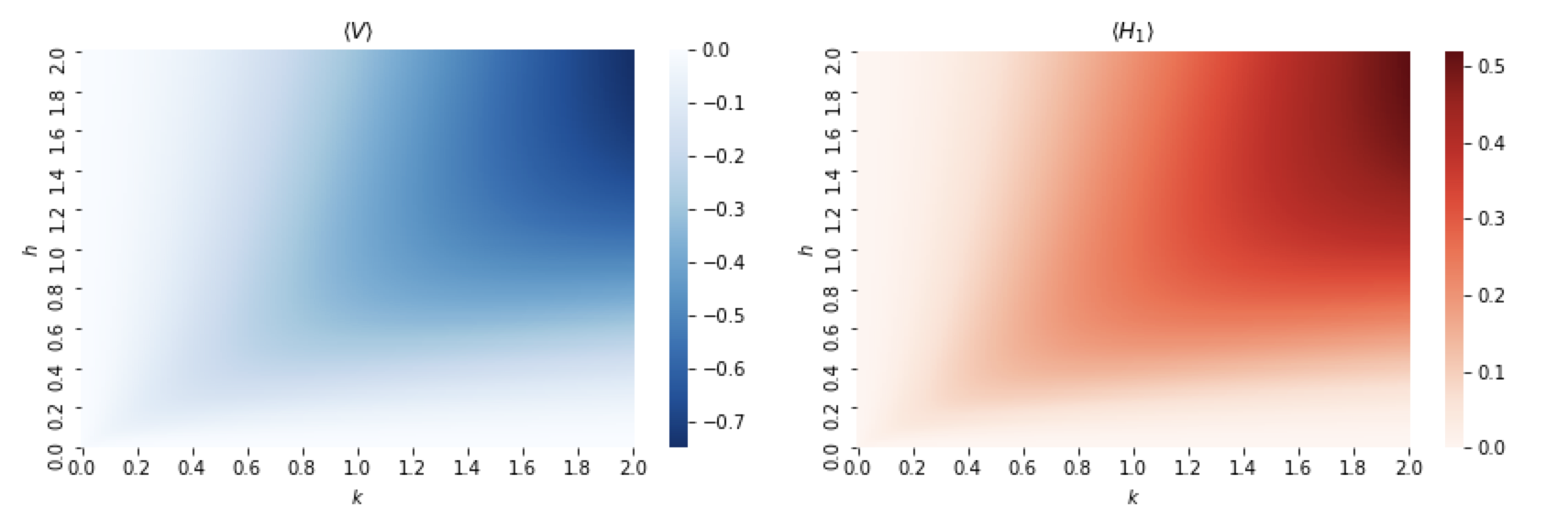}
    \caption{Heat maps visualizing expectation values $\langle V\rangle=\Tr[\rho_\text{QET}V]$ and $\langle H_1\rangle=\Tr[\rho_\text{QET}H_1]$ by $(k,h)$. }
    \label{fig:HV}
\end{figure*}
Here we describe the details of the model we used. For more information please refer to Hotta's original papers. First, it is important to note that the ground state of the total Hamiltonian $H$ is not the ground state of local operators. For example, $V$ has three degenerated ground states $\ket{-+},\ket{+-},\frac{\ket{-+}+\ket{+-}}{\sqrt{2}}$, and the ground state energy of $V$ is $-2k+\frac{2k^2}{\sqrt{h^2+k^2}}$. It is important that $V$'s ground state energy is negative for all $k>0$. This is also true for $H_{n}$, whose ground state energy is $-h+\frac{h^2}{\sqrt{h^2+k^2}}$. The expected values of $\langle V\rangle=\Tr[\rho_\text{QET}V]$ and $\langle H_1\rangle=\Tr[\rho_\text{QET}H_1]$ obtained by QET are shown in Fig.~\ref{fig:HV}.

To understand the non-triviality of the QET protocol, it is important to note that after Alice's measurement, no matter what unitary operation $W_1$ is performed on Bob's qubit, no energy can be extracted. This can be confirmed by 
\begin{equation}
\label{eq:Ene}
    \Tr[\rho_WH_\text{tot}]-\langle E_0\rangle=\bra{g}W^\dagger_1H_\text{tot}W_1\ket{g}\ge0,
\end{equation}
where
\begin{equation}
    \rho_W=W^\dagger_1\left(\sum_{\mu\in\{-1,1\}}P_0(\mu)\ket{g}\bra{g}P_0(\mu)\right)W_1.
\end{equation}
The inequality in eq.~\eqref{eq:Ene} is guaranteed by eq.~\eqref{eq:ground}.

If Bob does not perform any operations on his own system after Alice's measurement, the time evolution of Bob's local system is as follows
\begin{align}
\begin{aligned}
    \langle H_1(t)\rangle&=\Tr[\rho_Me^{itH}H_1e^{-itH}]=\frac{h^2(1-\cos(4kt))}{2\sqrt{h^2+k^2}}\\
    \langle V(t)\rangle&=\Tr[\rho_Me^{itH}Ve^{-itH}]=0, 
\end{aligned}
\end{align}
where $\rho_M=\sum_{\mu\in\{\pm1\}}P_0(\mu)\ket{g}\bra{g}P_0(\mu)$.

As a result of the natural time evolution of the system, energy is indeed transferred to Bob's local system, but this is no more than energy propagation in the usual sense. In QET, energy is not obtained through the natural time evolution of the system, but instantaneously as a result of communication. Since we consider a non-relativistic quantum many-body system, the speed of energy propagation is sufficiently slower than the speed of light. Classical communication, realized by optical communication, can convey information to remote locations much faster than the time evolution of physical systems. Hence, QET can be described as a fast energy propagation protocol.

It is known that the change in entropy before and after the measurement can be evaluated as follows
\begin{align}
    \Delta S_{AB}&=S_{AB}-\sum_{\mu\in\{\pm1\}}p_\mu S_{AB}(\mu)\\
    &\ge \frac{1+\sin^2\xi}{2\cos^3\xi}\ln\frac{1+\cos\xi}{1-\cos\xi}\frac{E_B}{\sqrt{h^2+k^2}}
\end{align}
where $p_\mu$ is the probability distribution of $\mu$,  $S_{AB}(\mu)$ is the entanglement entropy after the measurement, $\xi=\arctan\left(\frac{k}{h}\right)$ and $E_B$ is the amount of energy that Bob can receive ($E_B=-\langle E_1\rangle>0$)~\cite{2011arXiv1101.3954H}. Moreover the maximal energy that Bob would receive is bounded below by the difference of entropy:
\begin{equation}
    \max_{U_1(\mu)}E_B\ge \frac{2\sqrt{h^2+k^2}(\sqrt{4-3\cos^2\xi}-2+\cos^2\xi)\Delta S_{AB}}{(1+\cos\xi)\ln\left(\frac{2}{1+\cos\xi}\right)+(1-\cos\xi)\ln\left(\frac{2}{1-\cos\xi}\right)}.
\end{equation}

\section{\label{sec:Hotta}Simulation of Hotta's original QET protocol}
\begin{table*}
    \centering
    \begin{tabular}{|c|c|c|c|c|c|c|}\toprule
   $(h,k)$ &(1,0.1) &(1,0.2)& (1,0.5)& (1,1) &(1.5,1)\\\hline
   Analytical $\langle E_0\rangle$&0.9950 &0.9806&0.8944 & 0.7071& 1.2481\\
   \texttt{qasm\_simulator} $\langle E_0\rangle$& $0.9929\pm0.0010$ & $0.9807\pm0.0010$ & $0.8948\pm0.0010$ &$0.7067\pm0.0010$ &$1.2492\pm0.0015$\\
    Analytical $\langle V\rangle$&-0.0193 &-0.0701&-0.2598& -0.3746&-0.4905\\
    \texttt{qasm\_simulator} $\langle V\rangle$&$-0.0194\pm0.0057$ &$-0.0682\pm0.0011$&$-0.2625\pm0.0061$& $-0.3729\pm0.0063$ &$-0.4860\pm0.0061$\\
    Analytical $\langle H_1\rangle$ &0.0144&0.0521&0.1873& 0.2598&0.3480\\
    \texttt{qasm\_simulator} $\langle H_1\rangle$ &$0.0136\pm0.0006$& $0.0501\pm0.0011$&$0.1857\pm0.0022$& $0.2550\pm0.0028$ &$0.3493\pm0.0038$\\
    Analytical $\langle E_1\rangle$ &-0.0049&-0.0180& -0.0726& $-0.1147$ &-0.1425\\
    \texttt{qasm\_simulator} $\langle E_1\rangle$ & $-0.0058\pm0.0057$&$-0.0181\pm0.016$& $-0.0768\pm0.0064$ &$-0.1179\pm0.0068$&$-0.1367\pm0.0072$\\\hline
\end{tabular}
    \caption{Comparison between analytical values and numerical values from the quantum circuits with conditional operation~(Fig.~\ref{fig:operation} (B)). Each error corresponds to statistical error of $10^5$ shots. We evaluate $\langle E_1\rangle=\langle H_1\rangle+\langle V\rangle$.}
    \label{tab:simulator}
\end{table*}
Hotta's original QET protocol, which can be implemented by Fig.~\ref{fig:operation} (B) in the main text, does require the conditional operations based on a signal $\mu\in\{-1,+1\}$ that Bob receives from Alice. We performed quantum computation on the equivalent circuit (right quantum circuit in Fig.~\ref{fig:operation} (C)) that yielded exactly the same results. Let $\Lambda(U)=\ket{0}\bra{0}\otimes I+\ket{1}\bra{1}\otimes U$ be a controlled $U$ gate. Note that $\Lambda(U(-1))$ and $(X\otimes I)\Lambda(U(+1))(X\otimes I)$ commute:
\begin{figure}[H]
    \centering
    \includegraphics[width=\linewidth]{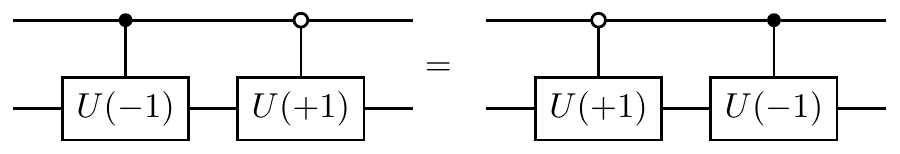}
\end{figure}

Of course, the equivalence of these circuits is theoretically trivial, we used \texttt{qasm\_simulator} and executed our simulation based on the left quantum circuit in Fig.~\ref{fig:operation} (C), in order to confirm the consistency between them. Table~\ref{tab:simulator} summarizes the numerical results and shows perfect agreement with the analytical results as well as results (Table~\ref{tab:complete}) with the right circuit in Fig.~\ref{fig:operation} (C).

\section{\label{sec:condition}Properties of Quantum Hardware}
\begin{figure}
    \centering
    \includegraphics[width=\linewidth]{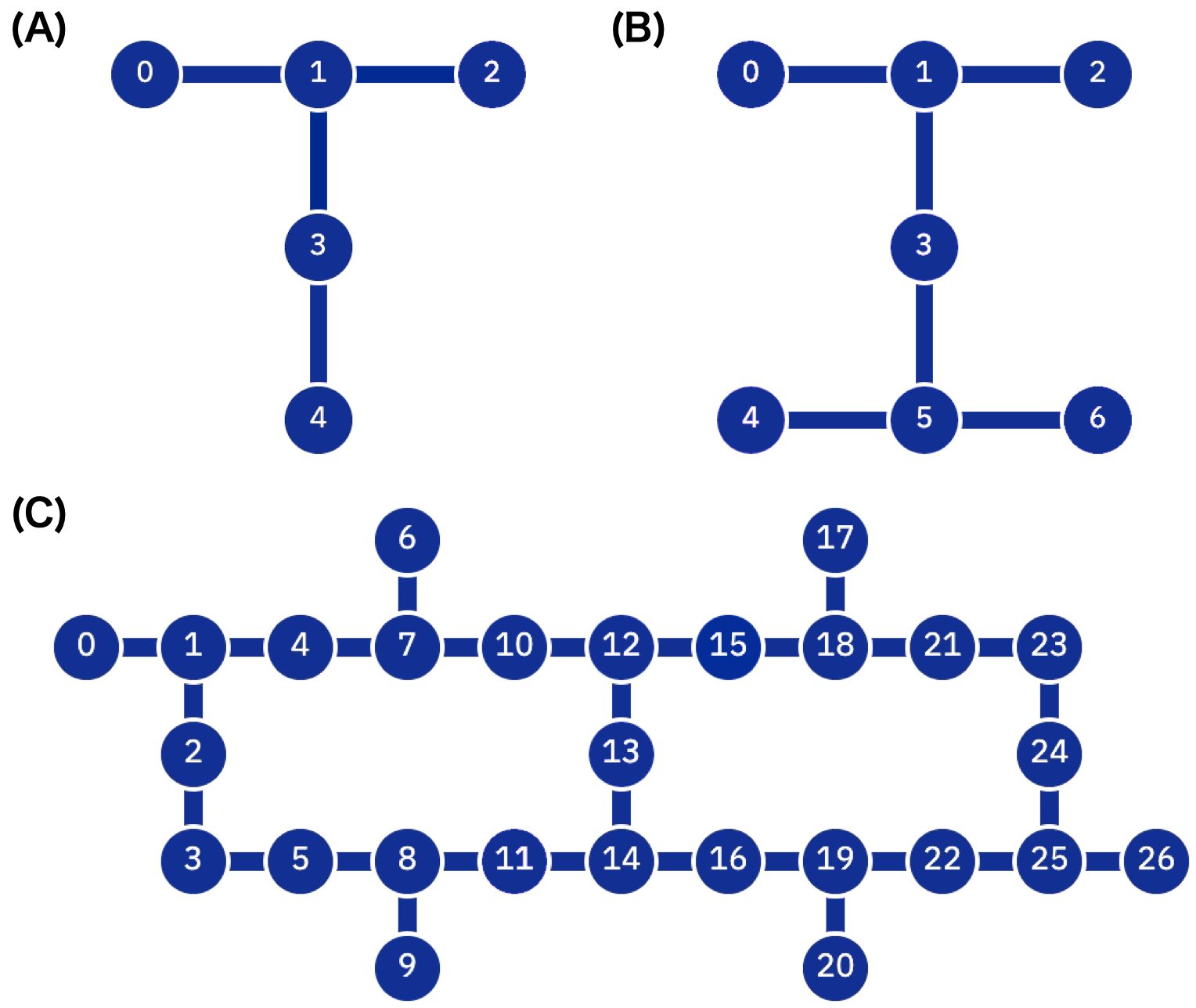}
    \caption{Configurations of qubits on graphs: (A) the layout of \texttt{ibmq\_lima} which has 5 qubits; (B) the layout of \texttt{ibmq\_jakarta} which has 7 qubits; (C) the layout of 27-qubit hardware including  \texttt{ibmq\_hanoi}, \texttt{ibm\_cairo}, \texttt{ibm\_auckland} and \texttt{ibmq\_montreal}. A direct CNOT gate can be applied to two qubits connected at the edge.}
    \label{fig:graph}
\end{figure}

Here we describe more on our experiments with 
IBM quantum computers. Graphs of IBM quantum computers we used are displayed in Fig~\ref{fig:graph}. For example, the layout of \texttt{ibmq\_lima} corresponds to (A) in Fig.~\ref{fig:graph} and we used the pair of qubits in [0,1] that had the smallest readout assignment error among all pairs (Fig.~\ref{fig:error_graph} (A) [Left]). We can perform a direct CNOT operation between qubits connected at the edge. For \texttt{ibmq\_lima}, the CNOT error between [1,2] qubits were 0.00510~(Table.~\ref{tab:processor}).

\begin{table*}
    \centering
    \begin{tabular}{|l|c|c|c|c|c|c|c|c|c|}\toprule
      \multicolumn{2}{|c|}{Backend} &\texttt{ibmq\_lima} &\texttt{ibmq\_jakarta} & \texttt{ibm\_cairo}  & \texttt{ibm\_hanoi}  &\texttt{ibmq\_auckland} &\texttt{ibmq\_montreal} \\\hline
       \multicolumn{2}{|c|}{$N_\text{tot}$}  & 5 & 7& 27 &27 &27&27\\
       \multicolumn{2}{|c|}{Quantum Volume} & 8 & 16& 64 &64 &64&128\\
       \multicolumn{2}{|c|}{shots}& $10^5$ & $3.2\times 10^4$ & $10^5$ &$10^5$ &$10^5$&$3.2\times 10^4$\\
       \multicolumn{2}{|c|}{Measurement fidelity}& 0.961075 & 0.924695 & 0.961935 &0.979530 &0.979383&0.957484\\
       \multicolumn{2}{|c|}{qubits used}& [0,1] & [3,5] & [13,14] &[14,16] &[14,16]&[14,16]\\
       \multicolumn{2}{|c|}{CNOT error}& 0.00510 & 0.00665 & 0.00439 &0.01996 &0.00570 & 0.00739\\
       \multicolumn{2}{|c|}{Gate time ($ns$)}& 305.778 & 291.556 & 220.444 &472.889 &355.556&355.556\\
       \multicolumn{2}{|c|}{CLOPS}& 2700 & 2400 & 2400 &2300 &2400&2000\\\hline
       \multirow{6}{*}{\begin{turn}{270}First qubit\end{turn}}&$t_1(\mu s)$& 75.67 & 93.53 & 146.43 &219.15 &60.97&129.56\\
       &$t_2(\mu s)$& 141.39& 41.09 & 164.29 &25.75 &150.49&168.53\\
       &Frequency (GHz)& 5.030 & 5.178 & 5.282 &5.047 &5.167&4.961\\
       &Anharmonicity (GHz)& -0.33574 & -0.34112 & -0.33874 &-0.34412 &-0.34196&-0.32314\\
       &Pauli $X$ error& $2.781\times10^{-4}$ & $2.140\times10^{-4}$ & $1.630\times10^{-4}$ &$2.305\times10^{-4}$ &$2.4842\times10^{-4}$&$1.942\times10^{-4}$\\
       &Readout assignment error& $1.960\times10^{-2}$ & $2.440\times10^{-2}$ & $8.500\times10^{-3}$ &$7.400\times10^{-3}$ &$8.100\times10^{-3}$&$1.310\times10^{-2}$\\\hline
       \multirow{6}{*}{\begin{turn}{270}Second qubit\end{turn}}&$t_1(\mu s)$& 58.03 & 143.52 & 94.28 &190.07 &73.16&83.73\\
       &$t_2(\mu s)$& 74.97 & 59.33 & 186.99 &253.46 &183.12 &39.92\\
       &Frequency (GHz)& 5.128 & 5.063 & 5.044 &4.883 &4.970&5.086\\
       &Anharmonicity (GHz)& -0.31835 & -0.34129 & -0.34289 &-0.34591&-0.34389&-0.33707\\
       &Pauli $X$ error& $1.469\times10^{-4}$ & $1.708\times10^{-4}$ & $1.732\times10^{-4}$ &$4.708\times10^{-4}$ &$2.052\times10^{-4}$&$2.221\times10^{-4}$\\
       &Readout assignment error& $1.300\times10^{-2}$ & $2.400\times10^{-2}$ & $8.000\times10^{-3}$ &$9.600\times10^{-3}$&$7.700\times10^{-3}$&$9.800\times10^{-3}$ \\
       \hline
    \end{tabular}
    \caption{Machine properties of IBM quantum computers and parameters we used. shots is the number of iterations we performed for sampling. Average measurement fidelity was computed when preparing a calibration matrix and used for measurement error mitigation. CNOT error corresponds to the direct CNOT error between two qubits $[q_0,q_1]$ used. Gate time corresponds to the gate time between $[q_0,q_1]$. The first and second qubits correspond to $q_0$ and $q_1$, respectively. $t_1$ is relaxation time and $t_2$ is dephasing time. For those who use \texttt{ibmq\_lima} with an open account, the maximal $n_\text{shots}$ is $3.2\times 10^4$. CLOPS means the number of circuit layer operations per second, indicating how many layers of a circuit a quantum processing unit (QPU) can execute per unit of time.}
    \label{tab:processor}
\end{table*}

\section{\label{sec:Add}Additional results with 6 different quantum hardware}
Here we describe additional results obtained by some other IBM quantum computers. In the main text we focused on best results with \texttt{ibmq\_lima} and \texttt{ibmq\_jakarta}, but in fact we also experimented with \texttt{ibmq\_hanoi}, \texttt{ibm\_cairo}, \texttt{ibm\_auckland}, \texttt{ibmq\_montreal}.
Table~\ref{tab:complete} summarizes the complete lists of the best data we obtained and Table~\ref{tab:processor} summarizes the experimental conditions used for each hardware. In the entire circuit, the total number $N$ of qubits is 2 and the circuit depth $d(N)$ that can be executed is 5 (excluding measurement of $V$) and 6 (including measurement of $V$).
The quantum volume is defined by $\text{QV}=\left(\text{arg max}_{n\le N}\min\{n,d(n)\}\right)^2$. Therefore quantum computers with $\text{QV}=64$ are enough for this work. Here QV is a metric that quantifies the largest random circuit of equal width and depth that a quantum computer can successfully implement. However, QV may not be a crucial metric in this study, since we are only dealing with 2-qubit, relatively simple quantum circuits. Errors in quantum computers result from a combination of various factors, including readout error, CNOT error, etc.. 
Table~\ref{tab:complete} shows that Alice's measurements of $X_0$ are relatively accurate in almost all cases. With respect to the observation of $V$, there is a deviation from the analytical value. It was confirmed that the error mitigation improved the results. In this study, what is important is that negative expectation values $\langle V\rangle$ were observed for all cases. It is a noteworthy achievement that negative energy expectation values $\langle E\rangle<0$ were observed by error mitigation. In fact, the histograms of states~(Fig.~\ref{fig:error_graph} (B)) have improved to approach the exact values, indicating that all operations were performed correctly.

\begin{widetext}
\begin{table}
    \centering
    \begin{tabular}{|c|c|c|c|c|c|}\toprule
\cline{1-6}
Backend & &$(h,k)=(1,0.2)$&$(h,k)=(1,0.5)$& $(h,k)=(1,1)$ &$(h,k)=(1.5,1)$ \\\cline{1-6}
Analytical value&$\langle E_0\rangle$&0.9806&0.894&0.7071&1.2481\\\hline
\multicolumn{1}{ |c  }{\multirow{2}{*}{\texttt{ibmq\_lima}} } &
\multicolumn{1}{ |c| }{error mitigated} &$0.9423\pm0.0032$& $0.8169\pm0.0032$& $0.6560\pm0.0031$ &$1.2480\pm0.0047$    \\ \cline{2-6}
\multicolumn{1}{ |c  }{}                        &
\multicolumn{1}{ |c| }{unmitigated} &$0.9049\pm0.0017$&$0.8550\pm0.0032$&$0.6874\pm0.0031$&$1.4066\pm0.0047$    \\\hline
\multicolumn{1}{ |c  }{\multirow{2}{*}{\texttt{ibmq\_jakarta}} } &
\multicolumn{1}{ |c| }{error mitigated} &$0.9299\pm0.0056$& $0.8888\pm0.0056$& $ 0.7039\pm0.0056$ &$1.2318\pm0.0084$ \\ \cline{2-6}
\multicolumn{1}{ |c  }{}                        &
\multicolumn{1}{ |c| }{unmitigated} &$0.9542\pm0.0056$&$0.9089\pm0.0056$&$ 0.7232\pm0.0056$&$1.2624\pm0.0083$\\ \cline{1-6}
\multicolumn{1}{ |c  }{\multirow{2}{*}{\texttt{ibm\_hanoi} } } &
\multicolumn{1}{ |c| }{error mitigated} &$1.0685\pm0.0032$&$0.9534\pm0.0032$& $0.7852\pm0.0031$ &$1.3728\pm0.0047$ \\ \cline{2-6}
\multicolumn{1}{ |c  }{}                        &
\multicolumn{1}{ |c| }{unmitigated}& $1.0670\pm0.0031$&$0.9524\pm0.0031$&$ 0.7809\pm0.0031$&$1.3663\pm0.0047$ \\\hline
\multicolumn{1}{ |c  }{\multirow{2}{*}{\texttt{ibm\_cairo}} } &
\multicolumn{1}{ |c| }{error mitigated} &$0.9571\pm0.0032$& $0.8626\pm0.0031$& $0.7277\pm0.0031$ &$1.2072\pm0.0047$    \\ \cline{2-6}
\multicolumn{1}{ |c  }{}                        &
\multicolumn{1}{ |c| }{unmitigated} &$0.9578\pm0.0031$&$0.8735\pm0.0031$&$0.7362\pm0.0031$&$1.2236\pm0.0047$    \\ \cline{1-6}
\multicolumn{1}{ |c  }{\multirow{2}{*}{\texttt{ibm\_auckland}} } &
\multicolumn{1}{ |c| }{error mitigated} &$0.9766\pm0.0032$& $0.8703\pm0.0032$& $0.6925\pm0.0032$ &$1.2482\pm0.0047$    \\ \cline{2-6}
\multicolumn{1}{ |c  }{}                        &
\multicolumn{1}{ |c| }{unmitigated} &$0.9771\pm0.0032$&$0.8712\pm0.0032$&$0.6931\pm0.0032$&$1.2487\pm0.0047$    \\ \cline{1-6}
\multicolumn{1}{ |c  }{\multirow{2}{*}{\texttt{ibmq\_montreal} } } &
\multicolumn{1}{ |c| }{error mitigated} &$0.8774\pm0.0056$&$0.8084\pm0.0056$& $0.6315\pm0.0056$ &$1.1449\pm0.0084$ \\ \cline{2-6}
\multicolumn{1}{ |c  }{}                        &
\multicolumn{1}{ |c| }{unmitigated} &$0.9036\pm0.0056$&$0.8338\pm0.0056$&$ 0.6564\pm0.0056$&$1.1819\pm0.0084$ \\ 
\toprule
Analytical value&$\langle H_1\rangle$&0.0521&0.1873&0.2598 &0.3480\\\hline
\multicolumn{1}{ |c  }{\multirow{2}{*}{\texttt{ibmq\_lima}} } &
\multicolumn{1}{ |c| }{error mitigated} &$0.0733\pm0.0032$& $0.1934\pm0.0032$& $0.2526\pm0.0032$ &$0.3590\pm0.0047$    \\ \cline{2-6}
\multicolumn{1}{ |c  }{}                        &
\multicolumn{1}{ |c| }{unmitigated} &$0.1295\pm0.0053$&$0.2422\pm0.0024$&$0.2949\pm0.0028$&$0.4302\pm0.0039$ \\ \cline{1-6}
\multicolumn{1}{ |c  }{\multirow{2}{*}{\texttt{ibmq\_jakarta}} }&
\multicolumn{1}{ |c| }{error mitigated} &$0.0736\pm0.0055$& $0.2018\pm0.0056$& $0.2491\pm0.0056$ &$0.3390\pm0.0084$ \\ \cline{2-6}
\multicolumn{1}{ |c  }{}                        &
\multicolumn{1}{ |c| }{unmitigated} &$0.0852\pm0.0022$&$0.2975\pm0.0045$&$ 0.3365\pm0.0052$&$0.4871\pm0.0073$\\ \cline{1-6}
\multicolumn{1}{ |c  }{\multirow{2}{*}{\texttt{ibm\_hanoi} } } &
\multicolumn{1}{ |c| }{error mitigated} &$0.1786\pm0.0032$&$0.3256\pm0.0032$& $0.4276\pm0.0032$ &$0.5890\pm0.0047$ \\ \cline{2-6}
\multicolumn{1}{ |c  }{}                        &
\multicolumn{1}{ |c| }{unmitigated} &$0.2012\pm0.0019$&$0.3427\pm0.0026$&$ 0.4378\pm0.0031$&$0.6104\pm0.0042$ \\\hline 
\multicolumn{1}{ |c  }{\multirow{2}{*}{\texttt{ibm\_cairo}} } &
\multicolumn{1}{ |c| }{error mitigated} &$0.0674\pm0.0032$& $0.1653\pm0.0031$& $0.2579\pm0.0031$ &$0.3559\pm0.0047$    \\ \cline{2-6}
\multicolumn{1}{ |c  }{}                        &
\multicolumn{1}{ |c| }{unmitigated} &$0.0905\pm0.0014$&$0.1825\pm0.0022$&$0.2630\pm0.0027$&$0.3737\pm0.0037$ \\ \cline{1-6}
\multicolumn{1}{ |c  }{\multirow{2}{*}{\texttt{ibm\_auckland}} } &
\multicolumn{1}{ |c| }{error mitigated} &$0.1218\pm0.0032$& $0.2004\pm0.0031$& $0.2181\pm0.0032$ &$0.3215\pm0.0047$    \\ \cline{2-6}
\multicolumn{1}{ |c  }{}                        &
\multicolumn{1}{ |c| }{unmitigated} &$0.1455\pm0.0017$&$0.2205\pm0.0023$&$0.2337\pm0.0027$&$0.3493\pm0.0038$    \\ \cline{1-6}
\multicolumn{1}{ |c  }{\multirow{2}{*}{\texttt{ibmq\_montreal} } } &
\multicolumn{1}{ |c| }{error mitigated} &$0.0897\pm0.0056$&$0.1618\pm0.0056$& $0.1921\pm0.0056$ &$0.2816\pm0.0084$ \\ \cline{2-6}
\multicolumn{1}{ |c  }{}                        &
\multicolumn{1}{ |c| }{unmitigated} &$0.1603\pm0.0032$&$0.2251\pm0.0041$&$ 0.2454\pm0.0049$&$0.3704\pm0.0068$ \\ 
\toprule
Analytical value&$\langle V\rangle$&-0.0701&-0.2598 &-0.3746 &-0.4905\\\hline
\multicolumn{1}{ |c  }{\multirow{2}{*}{\texttt{ibmq\_lima}} } &
\multicolumn{1}{ |c| }{error mitigated} &$-0.0655\pm0.0012$& $-0.2041\pm0.0031$& $-0.2744\pm0.0063$ &$-0.4091\pm0.0063$    \\ \cline{2-6}
\multicolumn{1}{ |c  }{}                        &
\multicolumn{1}{ |c| }{unmitigated} &$-0.0538\pm0.0011$&$-0.1471\pm0.0025$&$-0.1233\pm0.0041$&$-0.2737\pm0.0046$ \\ \cline{1-6}
\multicolumn{1}{ |c  }{\multirow{2}{*}{\texttt{ibmq\_jakarta}} } &
\multicolumn{1}{ |c| }{error mitigated} &$-0.0515\pm0.0022$& $-0.2348\pm0.0056$& $-0.3255\pm0.0112$ &$-0.4469\pm0.0112$ \\ \cline{2-6}
\multicolumn{1}{ |c  }{}                        &
\multicolumn{1}{ |c| }{unmitigated} &$-0.0338\pm0.0021$&$-0.1371\pm0.0046$&$ -0.0750\pm0.0075$&$-0.2229\pm0.0083$\\ \cline{1-6}
\multicolumn{1}{ |c  }{\multirow{2}{*}{\texttt{ibm\_hanoi}} } &
\multicolumn{1}{ |c| }{error mitigated} &$-0.1136\pm0.0013$&$-0.2820\pm0.0031$& $-0.3497\pm0.0063$ &$-0.5512\pm0.0063$ \\ \cline{2-6}
\multicolumn{1}{ |c  }{}                        &
\multicolumn{1}{ |c| }{unmitigated} &$-0.1061\pm0.0011$&$-0.2494\pm0.0022$&$ -0.2704\pm0.0034$&$-0.4767\pm0.0038$ \\\hline
\multicolumn{1}{ |c  }{\multirow{2}{*}{\texttt{ibm\_cairo}} } &
\multicolumn{1}{ |c| }{error mitigated} &$-0.0497\pm0.0013$& $-0.1968\pm0.0031$& $-0.2569\pm0.0063$ &$-0.3804\pm0.0063$    \\ \cline{2-6}
\multicolumn{1}{ |c  }{}                        &
\multicolumn{1}{ |c| }{unmitigated} &$-0.0471\pm0.0012$&$-0.1682\pm0.0026$&$-0.1733\pm0.0038$&$-0.3089\pm0.0045$    \\ \cline{1-6}
\multicolumn{1}{ |c  }{\multirow{2}{*}{\texttt{ibm\_auckland}} } &
\multicolumn{1}{ |c| }{error mitigated} &$-0.0138\pm0.0012$& $-0.0854\pm0.0032$& $-0.0591\pm0.0063$ &$-0.1887\pm0.0063$    \\ \cline{2-6}
\multicolumn{1}{ |c  }{}                        &
\multicolumn{1}{ |c| }{unmitigated} &$-0.0113\pm0.0012$&$-0.0665\pm0.0027$ &$-0.0046\pm0.0044$ &$-0.1412\pm0.0049$    \\ \cline{1-6}
\multicolumn{1}{ |c  }{\multirow{2}{*}{\texttt{ibmq\_montreal}} } &
\multicolumn{1}{ |c| }{error mitigated} &$-0.0157\pm0.0022$&$-0.1207\pm0.0056$& $-0.1275\pm0.0112$ &$-0.1967\pm0.0112$ \\ \cline{2-6}
\multicolumn{1}{ |c  }{}                        &
\multicolumn{1}{ |c| }{unmitigated} &$-0.0091\pm0.0021$&$-0.0764\pm0.0048$&$ -0.0043\pm0.0079$&$-0.0926\pm0.0089$ \\
\toprule
Analytical value&$\langle E_1\rangle$&-0.0180&-0.0726& -0.1147&-0.1425\\\hline
\multicolumn{1}{ |c  }{\multirow{2}{*}{\texttt{ibmq\_lima}} } &
\multicolumn{1}{ |c| }{error mitigated} &$0.0078\pm0.0034$& $-0.0107\pm0.0045$& $-0.0217\pm0.0071$ &$-0.0501\pm0.0079$    \\ \cline{2-6}
\multicolumn{1}{ |c  }{}                        &
\multicolumn{1}{ |c| }{unmitigated} &$0.0757\pm0.0054$&$0.0950\pm0.0035$&$0.1715\pm0.0050$&$0.1565\pm0.0060$ \\ \cline{1-6}
\multicolumn{1}{ |c  }{\multirow{2}{*}{\texttt{ibmq\_jakarta}} } &
\multicolumn{1}{ |c| }{error mitigated} &$0.0221\pm0.0059$& $-0.0330\pm0.0079$& $-0.0764\pm0.0125$ &$-0.1079\pm0.0140$ \\ \cline{2-6}
\multicolumn{1}{ |c  }{}                        &
\multicolumn{1}{ |c| }{unmitigated} &$0.0514\pm0.0030$&$0.1604\pm0.0064$&$ 0.2615\pm0.0091$&$0.2642\pm0.00111$\\ \cline{1-6}
\multicolumn{1}{ |c  }{\multirow{2}{*}{\texttt{ibm\_hanoi} } } &
\multicolumn{1}{ |c| }{error mitigated} &$0.065\pm0.0034$&$0.0436\pm0.0044$& $0.0779\pm0.0071$ &$1.2481\pm0.015$ \\ \cline{2-6}
\multicolumn{1}{ |c  }{}                        &
\multicolumn{1}{ |c| }{unmitigated} &$0.0950\pm0.0022$&$0.0933\pm0.0021$&$ 0.1674\pm0.0046$&$1.0566\pm0.015$ \\\hline
\multicolumn{1}{ |c  }{\multirow{2}{*}{\texttt{ibm\_cairo}} } &
\multicolumn{1}{ |c| }{error mitigated} &$0.0177\pm0.0035$& $-0.0315\pm0.0044$& $0.0010\pm0.0070$ &$-0.0245\pm0.0079$    \\ \cline{2-6}
\multicolumn{1}{ |c  }{}                        &
\multicolumn{1}{ |c| }{unmitigated} &$0.0433\pm0.0018$&$0.0143\pm0.0034$&$0.0897\pm0.0047$&$0.0648\pm0.0058$    \\ \cline{1-6}
\multicolumn{1}{ |c  }{\multirow{2}{*}{\texttt{ibm\_auckland}} } &
\multicolumn{1}{ |c| }{error mitigated} &$0.1080\pm0.0034$& $0.1149\pm0.0045$& $0.5877\pm0.0031$ &$1.2072\pm0.0047$    \\ \cline{2-6}
\multicolumn{1}{ |c  }{}                        &
\multicolumn{1}{ |c| }{unmitigated} &$0.1341\pm0.0021$&$0.154\pm0.0035$&$0.6364\pm0.0031$&$1.2236\pm0.0047$    \\ \cline{1-6}
\multicolumn{1}{ |c  }{\multirow{2}{*}{\texttt{ibmq\_montreal} } } &
\multicolumn{1}{ |c| }{error mitigated} &$0.0740\pm0.0060$&$0.0411\pm0.0079$& $0.0645\pm0.0057$ &$0.0849\pm0.0140$ \\ \cline{2-6}
\multicolumn{1}{ |c  }{}                        &
\multicolumn{1}{ |c| }{unmitigated} &$0.1512\pm0.0038$&$0.1487\pm0.0063$&$ 0.2411\pm0.0093$&$0.2778\pm0.0112$ \\ \cline{1-6}
\end{tabular}
    \caption{Results by \texttt{ibmq\_lima}, \texttt{ibmq\_jakarta}, \texttt{ibmq\_hanoi}, \texttt{ibm\_cairo}, \texttt{ibm\_auckland}, \texttt{ibmq\_montreal}.}
    \label{tab:complete}
\end{table}
\end{widetext}
\end{document}